\documentclass[manuscript]{acmart}
\usepackage{wrapfig}
\usepackage{url}
\usepackage{tikz}
\newcommand{\xmark}{%
\tikz[scale=0.23] {
    \draw[line width=0.7,line cap=round] (0,0) to [bend left=6] (1,1);
    \draw[line width=0.7,line cap=round] (0.2,0.95) to [bend right=3] (0.8,0.05);
}}
\newcommand{\cmark}{%
\tikz[scale=0.23] {
    \draw[line width=0.7,line cap=round] (0.25,0) to [bend left=10] (1,1);
    \draw[line width=0.8,line cap=round] (0,0.35) to [bend right=1] (0.23,0);
}}
\AtBeginDocument{%
  }
\usepackage{subcaption}
\setcopyright{acmcopyright}
\copyrightyear{2023}
\acmYear{2023}
\acmDOI{XXXXXXX.XXXXXXX}

\acmJournal{TOG}
\acmVolume{37}
\acmNumber{4}
\acmArticle{111}
\acmMonth{8}

\begin{document}

\title{Drifter: Efficient Online Feature Monitoring for Improved Data Integrity in Large-Scale Recommendation Systems}
\author{Bla\v{z} \v{S}krlj}
\email{bskrlj@outbrain.com}
\orcid{0000-0002-9916-8756}
\affiliation{%
  \institution{Outbrain Inc.}
  \city{Ljubljana}
  \country{Slovenia}
}

\author{Nir Ki-Tov}
\email{nktov@outbrain.com}
\affiliation{%
  \institution{Outbrain Inc.}
  \city{Netanya}
  \country{Israel}
}

\author{Lee Edelist}
\email{leedel2@gmail.com}
\affiliation{%
  \institution{-}
  \city{Netanya}
  \country{Israel}
}

\author{Natalia Silberstein}
\email{nsilberstein@outbrain.com}
\affiliation{%
  \institution{Outbrain Inc.}
  \city{Netanya}
  \country{Israel}
}

\author{Hila Weisman-Zohar}
\email{hila.weisman@gmail.com}
\affiliation{%
  \institution{-}
  \city{Netanya}
  \country{Israel}
}

\author{Bla\v{z} Mramor}
\email{bmramor@outbrain.com}
\affiliation{%
  \institution{Outbrain Inc.}
  \city{Ljubljana}
  \country{Slovenia}
}

\author{Davorin Kopi\v{c}}
\email{dkopic@outbrain.com}
\affiliation{%
  \institution{Outbrain Inc.}
  \city{Ljubljana}
  \country{Slovenia}
}

\author{Naama Ziporin}
\email{nziporin@outbrain.com}
\affiliation{%
  \institution{Outbrain Inc.}
  \city{Netanya}
  \country{Israel}
}

\renewcommand{\shortauthors}{\v{S}krlj et al.}

\begin{abstract}
Real-world production systems often grapple with maintaining data quality in large-scale, dynamic streams. We introduce Drifter, an efficient and lightweight system for online feature monitoring and verification in recommendation use cases. Drifter addresses limitations of existing methods by delivering agile, responsive, and adaptable data quality monitoring, enabling real-time root cause analysis, drift detection and insights into problematic production events. Integrating state-of-the-art online feature ranking for sparse data and anomaly detection ideas, Drifter is highly scalable and resource-efficient, requiring only two threads and less than a gigabyte of RAM per production deployments that handle millions of instances per minute (model training). Drifter's effectiveness in alerting and mitigating data quality issues was demonstrated on a real-life system that handles up to a billion predictions per second.
\end{abstract}




\maketitle

\section{Introduction}
\label{sec:intro}
Designing and developing online machine learning systems is a complex endeavour, where data quality and integrity plays a crucial role for models' online performance. This is in particular the case for contemporary recommender systems (see~\cite{hoi2021online}), which often rely on frequent model updates (every few minutes or even less) and are thus subject to especially rapid negative impact from data quality degradation. To address potential data quality issues, such recommender systems can considerably benefit from supporting data-monitoring systems that enable fast alerting/responsiveness when data-related issues are present.
Systems like Greykite~\cite{reza2021greykite-github} enable automated forecasting and facilitate profiling of emerging issues related to internal system behaviour. The study of online features' behavior is commonly referred to as \emph{online feature selection}. This branch of methods attempts to distill relevant features from irrelevant ones in an online learning setting~\cite{haug2020leveraging}. Approaches focusing on mining larger (online) data sets must be computationally efficient and easily interpretable~\cite{hoi2012online}. 
\textbf{Online feature selection} and ranking has also been a lively research endeavour for building real-time recommender systems. For example, ~\cite{wang2015online} demonstrated that feature groups are a possible way of efficient online feature selection since similar features tend to behave similarly in time. Furthermore, large amounts of data that require processing can already represent substantial computational burden as processing and transformation of instances can be expensive. It was shown that higher-dimensional feature spaces in online settings require specialized approaches that are versatile enough and can scale with real-life data sizes~\cite{manikandan2021feature}. 
The field of online feature ranking is a vibrant area of research and development; however, it needs to extend beyond the algorithmic aspects typically associated with it -- in addition to the algorithms, there is a growing need for systems that can effectively monitor features in \emph{real-time}. Design of such systems also encompasses implementing and deploying mechanisms that enable online inspection of feature scores and other related metrics (e.g., features' cardinalities and coverages).
In order to optimize utility and effectiveness, these monitoring systems cannot exist in isolation. They must be coupled with a visibility layer incorporating \emph{alerting mechanisms} and visualizations. This integration allows for an accessible and user-friendly interface, enabling the study of granular details of the data consumed by real-time models. By providing such visibility, users can gain insights into the underlying factors influencing the performance and behaviour of these models, and make informed decisions and analysis based on them.

The work presented in this paper builds upon recent ideas and advances in both online feature ranking and real-time signal analysis. Drawing on these domains, we developed a system that has been deployed in a \textbf{large-scale online cloud environment}. This system handles real-life data streams for various use cases, including click-through rate prediction, conversion rate prediction, and item viewability prediction. Furthermore, by operating in real-world and real-time settings, the system is actively used when handling complex, dynamic data monitoring scenarios.
Overall, the paper emphasizes the importance of online feature monitoring and highlights the value of integrating this functionality with a comprehensive visibility layer. Through this integration, the presented system offers a powerful tool for inspecting and analyzing the most granular levels of data used in real-time models, ultimately contributing to improved decision-making and performance optimization in various application domains.

The remainder of this paper is structured as follows. We begin by discussing the existing use cases where online feature monitoring and verification was used to facilitate deployment, monitoring and understanding of online recommender systems. Second, we describe Drifter, the system used for online feature monitoring and verification, its implementation, and the metrics implemented for measuring feature drifts and anomalies online. Third, we describe a use case where Drifter helped profile and identify features that were subject to drift. Finally, we discuss the implications and lessons learned when deploying and designing Drifter.

\section{Online feature monitoring overview}
\label{sec:rationale}
We describe the main use cases where online feature monitoring is a suitable approach for understanding, mitigating and improving online learning processes.

\textbf{Introduction of new features}
A typical process in many online learning workflows involves the introduction of new signals or features. However, due to the dynamic nature of online learning, incorporating new features often requires substantial engineering efforts that may span multiple teams. Unfortunately, various factors, such as miscommunications, logging bugs, or other issues can unintentionally impact these features' distribution, coverage, or relative importance.
To mitigate these challenges, it is crucial to have a mechanism in place to monitor and automatically raise alerts based on predefined conditions that indicate problematic change in feature values. By proactively detecting and addressing such issues as early as possible in the data pipeline, we are able to prevent the deployment of models that could negatively affect the business.
Furthermore, understanding how existing features vary in time provides an additional perspective on their behaviour. This knowledge can be valuable for prioritizing testing of new features and their transformations, used by the predictive model.

\textbf{Measuring quality drops of existing features}
\label{sec-measuring-quality}
As soon as an online feature monitoring system is capable of running in real-time, it can (and does) serve as a "ghost mode" for a given data stream (consumed by, e.g., click-through rate or conversion rate prediction models). Furthermore, by systematically measuring features' properties such as cardinality, coverage, and statistics such as quantiles and histograms, the monitoring system can, in a matter of milliseconds, alert relevant model stakeholders that a change in the distribution of an existing feature has occurred. Such events can occur due to multiple reasons; examples include changes in a component that participates in the construction or final transformation of the feature, a drop of data quality due to an external event and feature drifts -- gradual changes in a feature's distribution, eventually resulting in problematic behaviour~\cite{barddal2017survey,barddal2015analyzing}.

\textbf{Debugging online models}
\label{sec-debugging-online}
When working with data streams in online learning systems, explainability can become a challenge -- the use of deep factorization machine-based models or similar variants have made it increasingly difficult and time-consuming to inspect problematic models directly. However, by establishing a connection between a model's behaviour and detectable shifts in feature distributions, it becomes possible to conduct more systematic \emph{post-hoc} evaluations of the model itself.
One such example includes perturbation-based analysis, which focuses on identifying the effects of the distribution shifts of single or multiple features. Observing such shifts automatically online makes it easier to link them to the model's behaviour and gain insights into its performance and potential issues. Further, understanding temporal behaviour facilitates the design of follow-up offline experiments that help identify potentially useful transformations of existing features. Understanding feature distribution shifts thus enables a more structured evaluation of the model, facilitating the identification and resolution of problems that may arise during online learning.

\textbf{Understanding of temporal dynamics of features}
\label{sec-temporal-dynamics}
The endeavour to study the behaviour of multiple features online simultaneously is not necessarily considered due to its time-consuming nature. However, by being able to observe whole feature spaces' dynamics in time, patterns related to the complementary nature of features can arise, deepening the data scientists' understanding of which features fluctuate together; understanding temporal relations can help with the creation of new features that account for this dynamics, or facilitate exploration of alternative features that would otherwise be ignored. For example, separate teams can introduce features in isolation, not being aware of their complementary nature -- by visualizing the joint space, such patterns can be studied and can further simplify existing models.

\textbf{Online ranking of features' contributions}
\label{sec-online-ranking}
Productization of new signals (features) can be expensive and time-consuming. By simulating a feature's behaviour with the target space of interest, prioritization of new features can be facilitated, saving valuable resources that would otherwise be spent on multiple deployments, running A/B tests and other costly procedures, that were always inevitable for testing the new features' contribution to the target space. We proceed with an overview of \textbf{related work}.

We continue the discussion with an overview of existing feature monitoring/inspection systems and how they compare to Drifter. An overview of how Drifter compares to existing products and tools is summarized in Table~\ref{tbl:ref-comp}. The selected tools include existing, well-established solutions for online feature store construction and subsequent machine learning, as well as up-an-coming solutions.
\begin{table}[htb!]
\caption{Overview of existing online feature monitoring methods and their main (out-of-the-box) properties.}
\resizebox{\textwidth}{!}{

\begin{tabular}{l|lllllllll}
method                   & Own ML engine & Feature constr. & Extensions & Streaming statistics & Sparse data & Metrics support & Drift detection & Online visualiz. & Main use case                                                \\ \hline
feathr (LinkedIn)        & \xmark                       & \cmark                  & \cmark        & --                    & --              & \cmark           & \xmark              & \cmark                          & Online store    \\
hopsworks                & \xmark                       & \cmark                  & \cmark        & \xmark                         & --              & \cmark           & \xmark              & --                      & End-to-end platform                   \\
RasgoQL                  & \xmark                       & --              & \xmark         & \xmark                         & --              & \xmark            & \xmark              & \xmark                           & Scaling Pandas                        \\
Vertex AI                & \cmark                      & \cmark                  & \cmark        & \cmark                        & \cmark                  & --       & \cmark             & --                      & End-to-end platform                   \\
Tecton                   & \xmark                       & \cmark                  & \cmark        & \xmark                         & \cmark                  & \cmark           & \xmark              & --                      & End-to-end platform                   \\
Amazon SAGEMAKER Store   & \xmark                       & --              & \cmark        & --                         & \cmark                  & \cmark           & \xmark              & --                      & End-to-end platform                   \\
butterfree               & \xmark                       & \cmark                  & --    & \xmark                         & \cmark                  & \xmark            & \xmark              & \xmark                           & End-to-end platform                   \\
ByteHub\footnote{test}                  & \xmark                       & \cmark                  & --    & \xmark                         & --              & \xmark            & \xmark              & \xmark                           & Online store    \\
Databricks feature store & \xmark                       & \cmark                  & \cmark        & --                    & --              & --       & \xmark              & --                      & Online store    \\
Drifter (this paper)     & \cmark                      & \cmark                  & --    & \cmark                        & \cmark                  & \cmark           & \cmark             & \cmark                          & Drift detection and alerting
\end{tabular}
}
\label{tbl:ref-comp}
\end{table}
The three possible table marks represent complete feature/capability ($\cmark$), lack thereof ($\xmark$) or partial compatibility (--). The categories of comparison were selected in a way to entail different properties of this type of systems; from their extendability, compliance with sparse data formats to less known functionalities such as drift detection and support for computation of streaming statistics (sketching algorithms). The more extensive solutions include feathr\footnote{\url{https://github.com/feathr-ai/feathr}}, Vertex AI\footnote{\url{https://cloud.google.com/vertex-ai/docs/featurestore}}, SAGEMAKER Store\footnote{\url{https://aws.amazon.com/sagemaker/feature-store}}, Tecton\footnote{\url{https://www.tecton.ai/}} and Databricks feature store\footnote{\url{https://docs.gcp.databricks.com/machine-learning/feature-store/index.html}}. Other considered systems include ByteHub\footnote{\url{https://github.com/bytehub-ai/bytehub}}, butterfree\footnote{\url{https://github.com/quintoandar/butterfree}}, RasgoQL\footnote{\url{https://github.com/rasgointelligence/RasgoQL}} and hospworks\footnote{\url{https://github.com/logicalclocks/hopsworks}}. The capabilities were assessed based on products' websites and readily available documentation. We observed that most end-to-end tools could be, with varying degrees of effort, extended to fulfuil missing/partially missing fields of comparison (we evaluated out-of-the-box capabilities that require minimum engineering overhead). Finally, systems that do not employ their own engines are mostly based around Scikit-learn library~\cite{pedregosa2011scikit}.

\section{Drifter - an overview}
\label{sec:overview}
Having established the motivating use cases which led us to build Drifter, we continue with its overview, design and implementation choices, user interaction with the service and future applications.

\subsection{Overall architecture and implementation}
\label{sec-arch}
\begin{figure*}[t!]
     \centering
      \Description[Visualization of anomaly plots.]{}
     \begin{subfigure}[b]{0.8\textwidth}
         \centering
         \includegraphics[width=\textwidth, height=7cm]{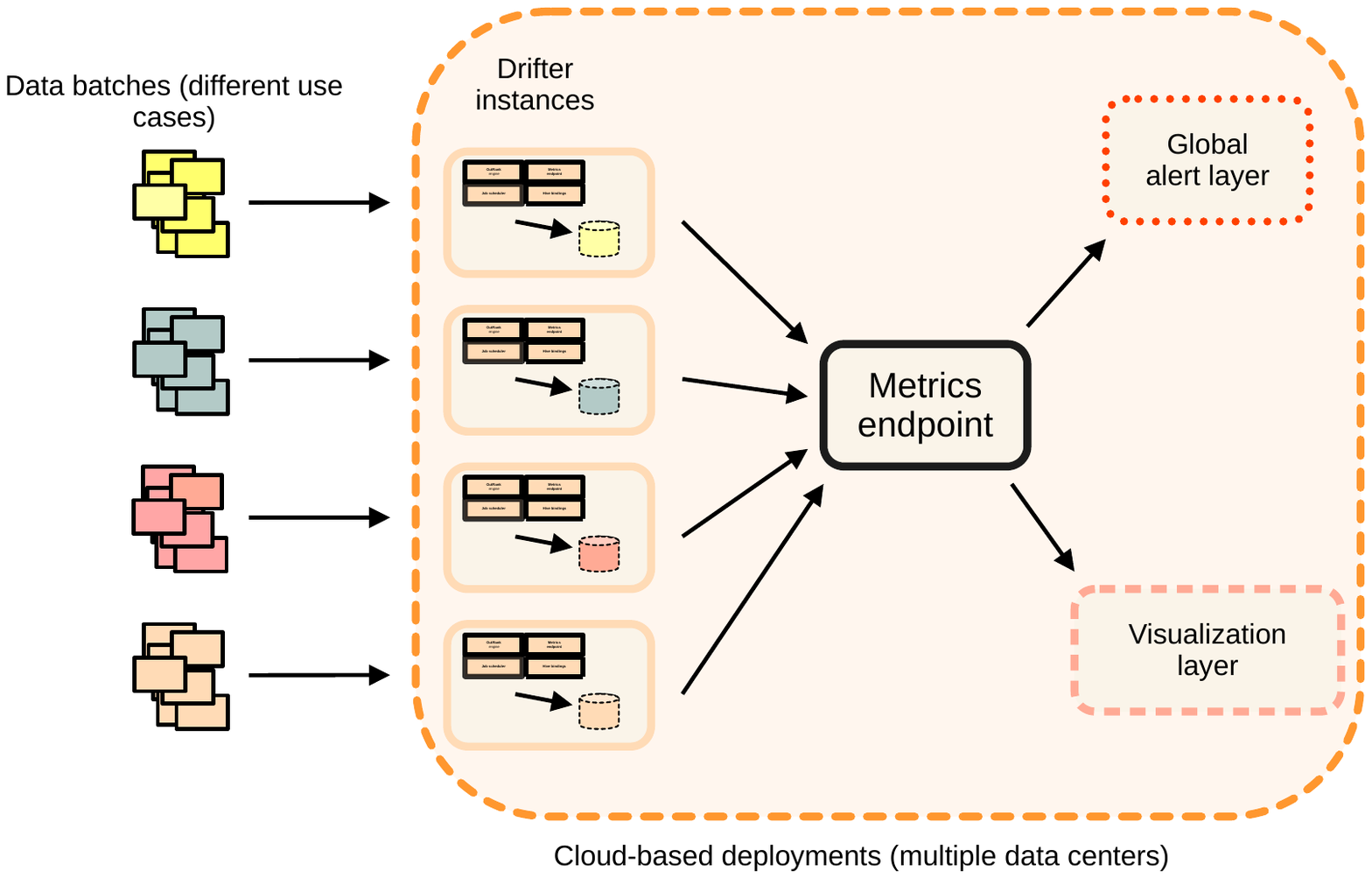}
         \caption{Cloud-based perspective of the Drifter service.     \vspace{.5cm}}
         \label{fig:drifter-single}
     \end{subfigure}
     \begin{subfigure}[b]{0.8\textwidth}
         \centering
         \includegraphics[width=\textwidth, height=7cm]{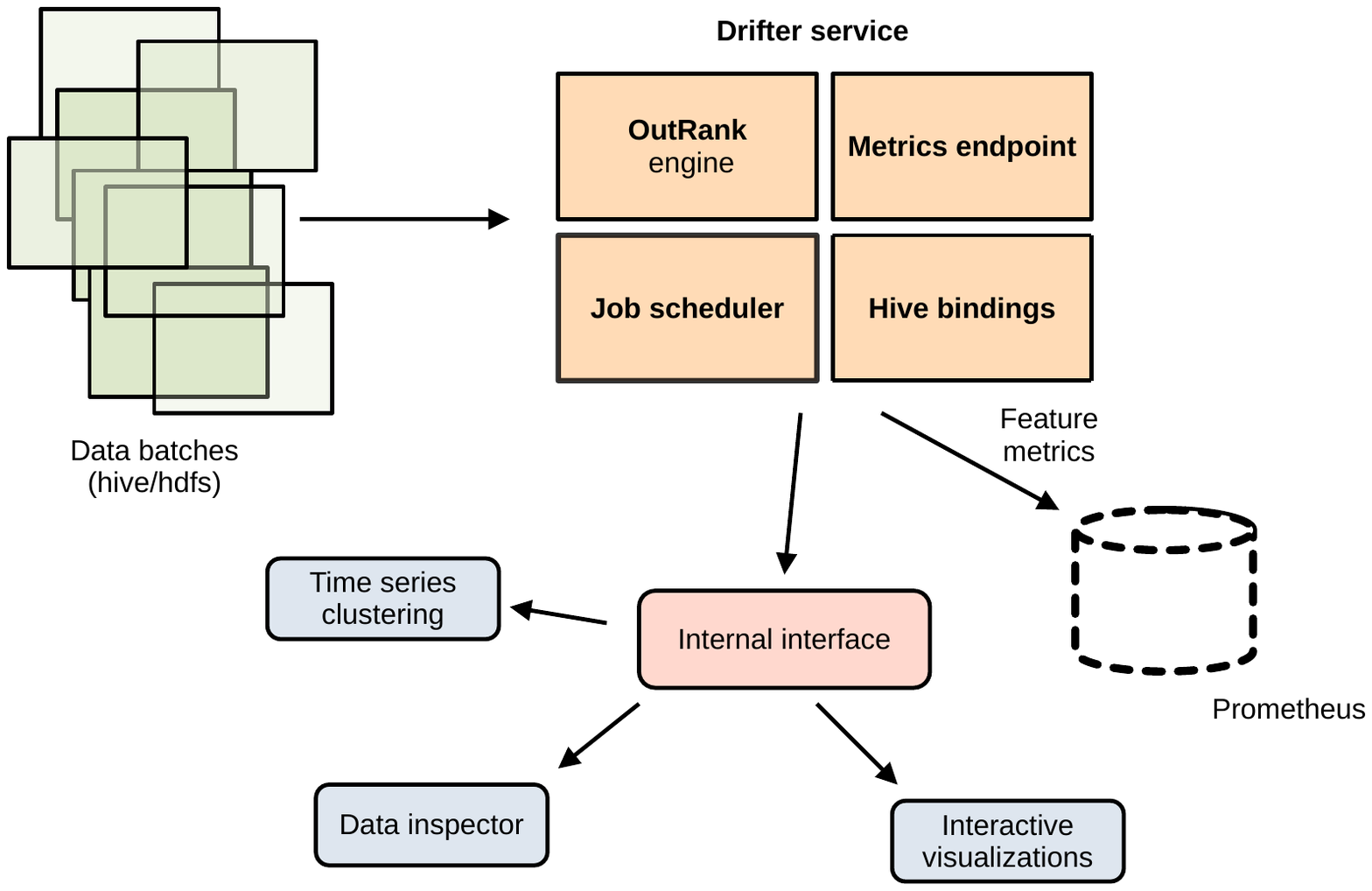}
         \caption{Overview of a single Drifter service instance.}
         \label{fig:drifter-cloud}
     \end{subfigure}
        \caption{Overview of Drifter and its placement into a broader (cloud computing grid) context. (a) The second layer of services is comprised of individual deployments (Figure 1). Their metrics are jointly aggregated, enabling the infrastructure team to monitor the global landscape of use cases for potentially extreme events (e.g., missing feature values, data leaks or similar). (b) Each instance has a scheduling system that governs the data updates/ingestion. The ranking engine enables fast feature profiling that can be inspected at the level of each Drifter pod or in Grafana-based dashboards derived by the produced metrics.}
        \label{fig:overivew}
\end{figure*}

Drifter was built as a component of the \textbf{microservice architecture}. It was built to operate with a distributed data source and a metrics endpoint of choice; in the presented work, however, it is a stand-alone service that receives the data via Hive\footnote{\url{https://hive.apache.org/}}, and outputs the metrics to a metrics service (Prometheus\footnote{\url{https://prometheus.io/}}). This design choice was undertaken so that each use case (e.g., a team owning a CTR prediction  or a viewability prediction model) has ownership over the relevant Drifter instance(s), and can modify their queries or data regime  according to their preferences. Each metrics endpoint is aware of a particular deployment, meaning that Grafana dashboards can be built with specific Drifter instance(s) in mind. This way, isolation of metrics is possible, but at the same time, teams can use other teams' metrics and information when defining their visualizations and alerts. The service itself is deployed to an in-house cloud platform, where each Drifter instance is monitored by default (resource-wise), offering users insights into the amount of resources required per each Drifter pod -- this functionality is useful for profiling changes in data regimes and their impact on the resources (which are finite for each use case). A single Drifter instance is summarized in Figure~\ref{fig:drifter-single}. 
Each Drifter is a self-contained unit that can be deployed on a per-demand basis in the in-house cloud platform. Although multiple different Drifters are simultaneously deployed (different use cases), their metrics are aggregated in a joint endpoint (Prometheus), enabling a global overview of features being monitored and their states. This overview is shown in Figure~\ref{fig:drifter-cloud}.

We continue with a more detailed overview of each of the components that constitute Drifter (service). Drifter is implemented as a Python-based service, utilizing an in-house library that, out-of-the-box, enables reporting of Prometheus-based metrics per deployed pod. Each Drifter has its internal scheduler, enabling flexibility in terms of time zones. In the initial phase of development, we observed two main computational bottlenecks to running Drifter instances online at scale: Consuming production-level data volumes comprised of up to millions of instances per ten minutes, and computing scores between features of interest. The computationally expensive parts of feature ranking related to score computing are written in Numba~\cite{lam2015numba} (an LLVM-based Python JIT compiler). Dockerized Drifter instances communicate with Prometheus endpoints (metrics). A Grafana-based dashboard enables the inspection of these metrics in real-time. The service is implemented with ease of on-boarding in mind; when a new use case needs to be accommodated (e.g., a novel model), a template Drifter pod is cloned and configured to operate with a dedicated data stream specific to a given use case. This way, use cases are separate and do not interfere with one another. Further, as they jointly push metrics to the common endpoint, visibility at the level of all active Drifter pods is possible and facilitates monitoring of their health by the infrastructure team.

\begin{figure*}[t!]
     \centering
      \Description[Visualization of anomaly plots.]{}
     \begin{subfigure}[b]{0.9\textwidth}
         \centering
         \fbox{\includegraphics[width=\linewidth, height=5cm]{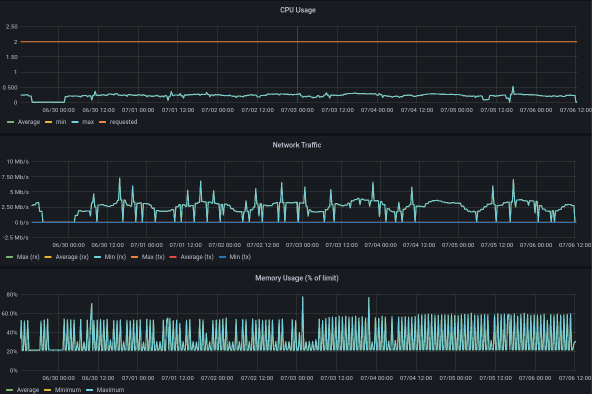}}
         \caption{Overview of resource utilization of Drifter instance (seven day period) that monitors CTR-related traffic. Drifter pod is stable even during traffic peaks (last sub-plot). CPU utilization is minimal (first plot). \vspace{0.3cm}}
         \label{fig:ctr-drift}
     \end{subfigure}
     \begin{subfigure}[b]{0.9\textwidth}
         \centering
         \fbox{\includegraphics[width=.7\textwidth, height=5cm]{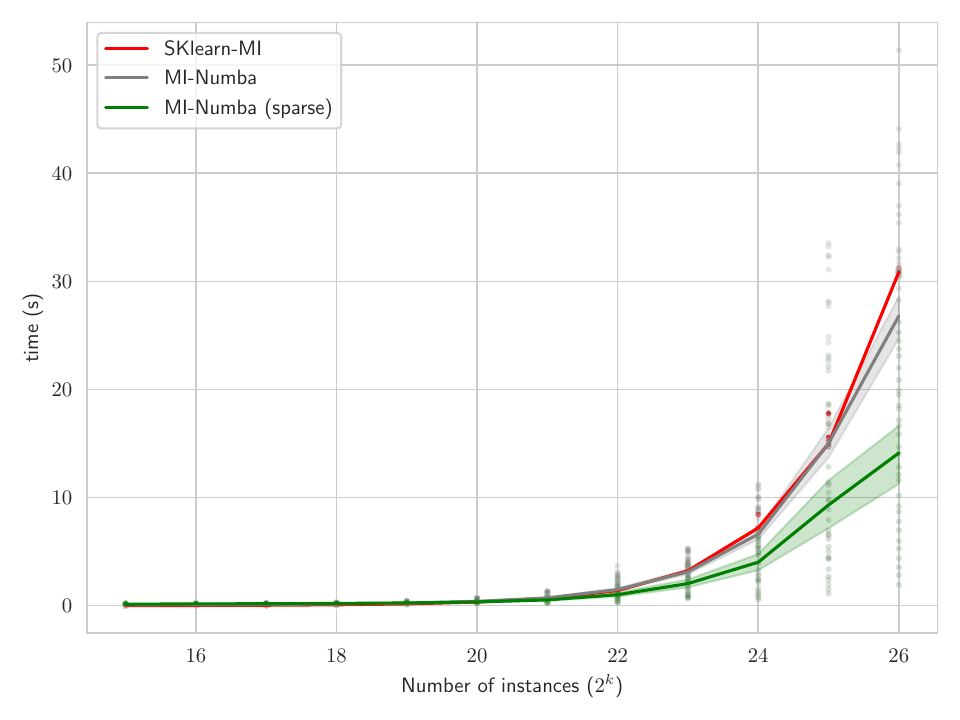}}
         \caption{Performance of Numba-based mutual information implementation against the established Scikit-learn one (different configurations ran ten times). Green samples represent a realistic scenario where only 30\% or less values are present.}
         \label{fig:ctr-numba}
     \end{subfigure}
        \caption{One colour represents one distinct feature. For this model, daily fluctuations in coverage are apparent for some features (e.g., the pink-coloured trace at the bottom part).}
        \label{fig:global-ctr-benchmark}
\end{figure*}

In order for Drifter to be available to as many use cases as possible, we optimized the service to a point it requires \textbf{less than 1GB of RAM and only two threads}. This was achieved by inspecting and optimizing Hive queries and the ranking engine itself. Optimizations that enable such low footprint include mini-batch feature ranking, probabilistic estimation of cardinalities (Hyperloglog-based counting), hashing trick for clipping values to a fixed (integer) range and randomized estimation of feature interactions -- for each mini-batch, the number of interactions computed is upper-bounded by a fixed maximum number, ensuring consistent performance. Overview of a live benchmark of the service on a week of production data is shown in Figure~\ref{fig:global-ctr-benchmark}. Memory limit (last plot in (a)) was set to 1GB. It can be observed that on average less than a single CPU is used. Memory spikes observed around e.g., 3rd of July correlate to variability in data quantity received by the service, showing resilience of Drifters to traffic spikes and similar events. Further, it can be observed that fluctuations in data have an impact that is within the resource constraints of the deployed pod. Each Drifter instance, as soon as it's deployed, produces metrics for resource consumption (apart from the ones related to feature space). Subfigure (b) demonstrates performance if the in-built ranking engine, in particular Numba-based re-implementation of mutual information. The algorithm was further extended to skip computations that are redundant due to data sparsity - benchmark shows evaluations of input vectors of different sizes with varying degrees if in-built sparseness (from 1\% to 50\% present values). For very sparse inputs, sparsity-aware mutual information can be substantially faster (on average it is comparable to the baseline).

\subsection{Visualization layer}

\begin{figure*}[b!]
     \centering
      \Description[Visualization of anomaly plots.]{}
     \begin{subfigure}[b]{0.85\textwidth}
         \centering
         \fbox{\includegraphics[width=\textwidth, height=5cm]{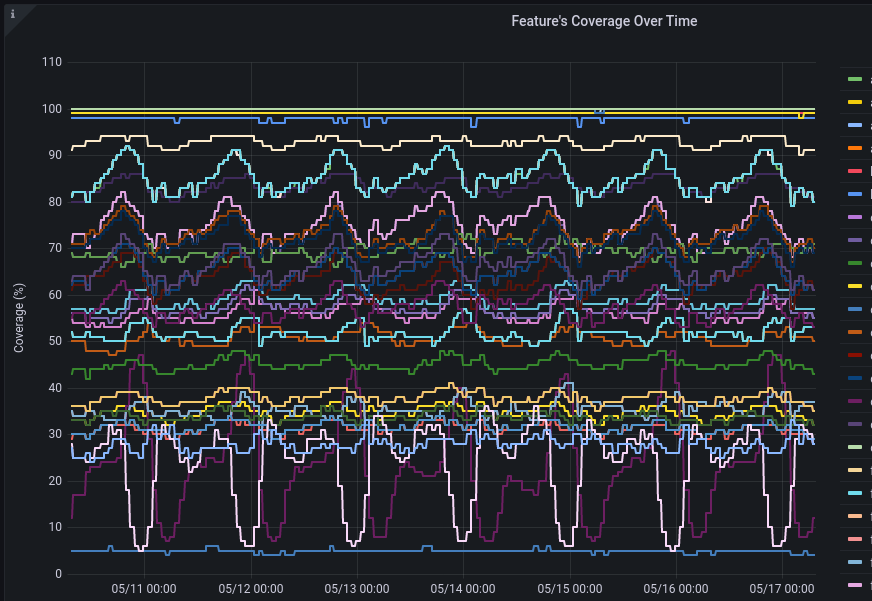}}
         \caption{One week of coverage information for a given feature space (click-through rate model).\vspace{0.5cm}}
         \label{fig:ctr-coverages}
     \end{subfigure}
     \hfill
     \begin{subfigure}[b]{0.85\textwidth}
         \centering
         \fbox{\includegraphics[width=\textwidth, height=5cm]{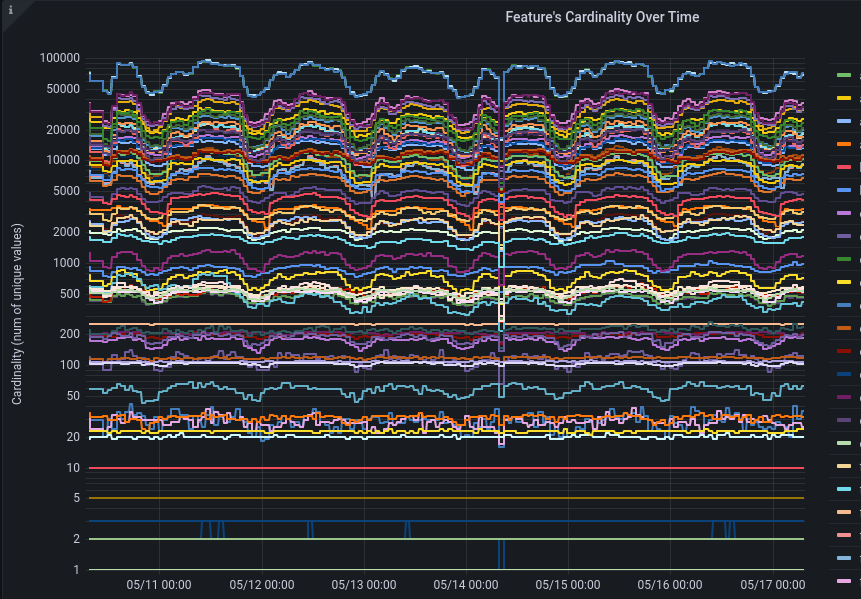}}
         \caption{Cardinality fluctuations in time.}
         \label{fig:ctr-cardinalities}
     \end{subfigure}
        \caption{One colour represents one distinct feature. For this model, daily fluctuations in coverage are apparent for some features (e.g., the pink-coloured trace at the bottom part).}
        \label{fig:overivew}
\end{figure*}

A vital component of each Drifter instance are its resulting visualizations. The design choice of metric-based visualizations (Prometheus and Grafana) enabled us to generalize metric retrieval and, at the same time, facilitate the creation of custom metrics and alerts based on them. Further, as Prometheus comes equipped with a collection of aggregation functions, users can create complex queries based on many existing examples while also sharing the knowledge -- resulting PromQL queries can easily be shared and tested across use cases. Similarly, Grafana-based visualizations are a direct extension of the metrics. Out-of-the-box capabilities suffice for most use cases, are flexible and customizable, and are easy to maintain or change if needed. An example visualization of feature coverage in time is shown in Figure~\ref{fig:ctr-coverages}.
A similar view showing features' cardinalities is shown in Figure~\ref{fig:ctr-cardinalities}.

A more complex example includes the computation of drifts -- changes in the feature's value within a given (parametrized) time frame. Parameterization of this aspect was necessary, as different use cases adhere to different temporal dynamics. An example is shown in Figure~\ref{fig:overivew}.

\begin{figure*}[h]
     \centering
      \Description[Visualization of anomaly plots.]{}
     \begin{subfigure}[h]{0.85\textwidth}
         \centering
         \fbox{\includegraphics[width=\textwidth, height=5cm]{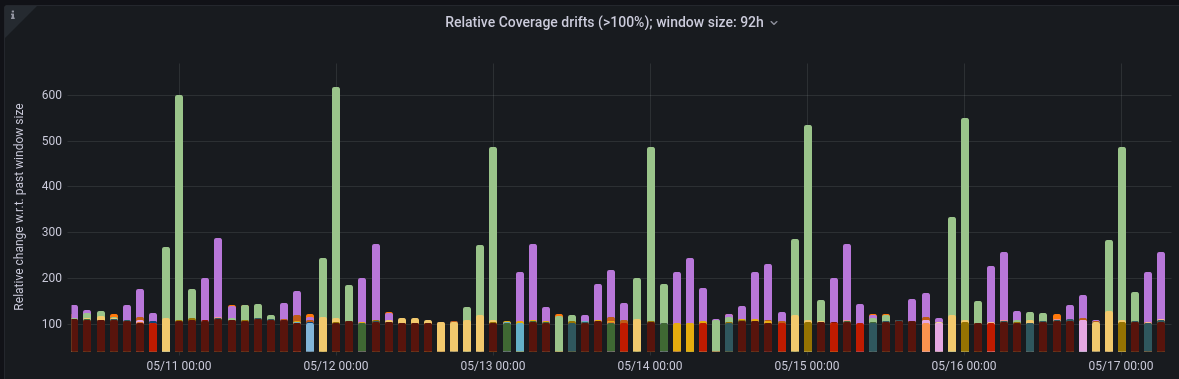}}
         \caption{An example visualization of a metric associated with feature drifts. Periodic and anomalous behavior are considered.\vspace{0.5cm}}
         \label{fig:ctr-drifts}
     \end{subfigure}
     \hfill
     \begin{subfigure}[h]{0.85\textwidth}
         \centering
         \fbox{\includegraphics[width=\textwidth, height=5cm]{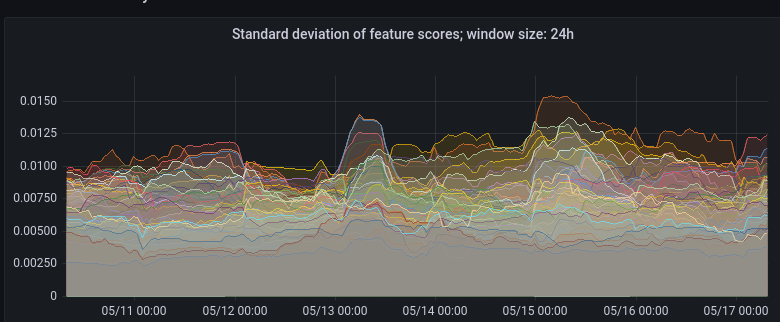}}
         \caption{Apart from basic statistics such as coverage and cardinality, Drifter also performs \textbf{online feature ranking}.}
         \label{fig:ctr-ranks}
     \end{subfigure}
        \caption{Overview of online feature drift-related metrics. (a) Large coverage changes are problematic and can be directly detected by Drifter. Every colour represents a unique feature. The green (highest) bars indicate a feature that varies in coverage (in time).}
        \label{fig:overivew}
\end{figure*}
The visualizations above are possible with out-of-the-box Grafana capabilities. However, more custom visualizations are also possible and are implemented at the pod level. An example includes interactive hierarchical clusters of time series of features' cardinalities.

\section{Use case example: Anomaly detection}
\label{sec-usecase}
Having discussed the overall Drifter architecture, we proceed with a collection of production use cases where Drifter was used to facilitate and enable data-related monitoring.

\begin{figure*}[h!]
     \centering
      \Description[Visualization of anomaly plots.]{}
          \begin{subfigure}[b]{0.85\textwidth}
         \centering
         \fbox{\includegraphics[width=\textwidth, height=5cm]{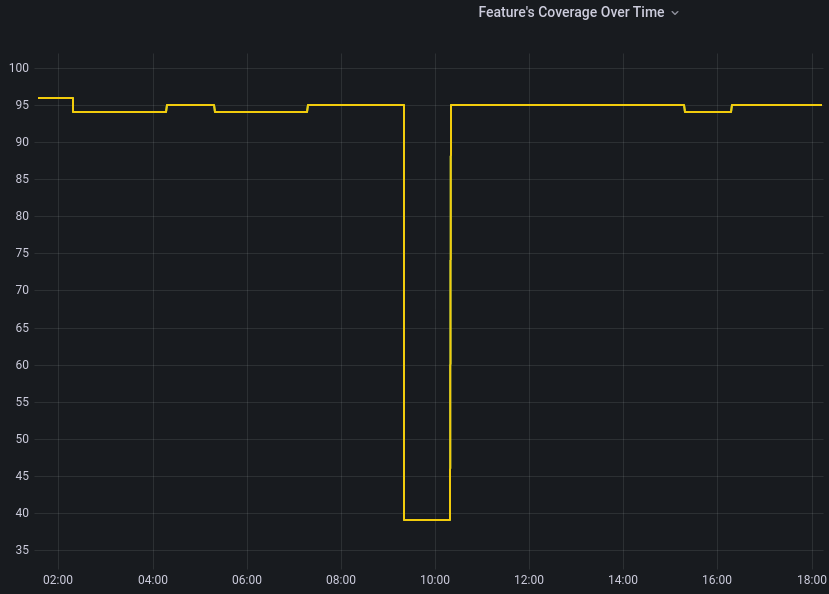}}
         \caption{A feature's coverage.}
     \end{subfigure}
     \hfill
     \begin{subfigure}[b]{0.85\textwidth}
         \centering
         \fbox{\includegraphics[width=\textwidth, height=5cm]{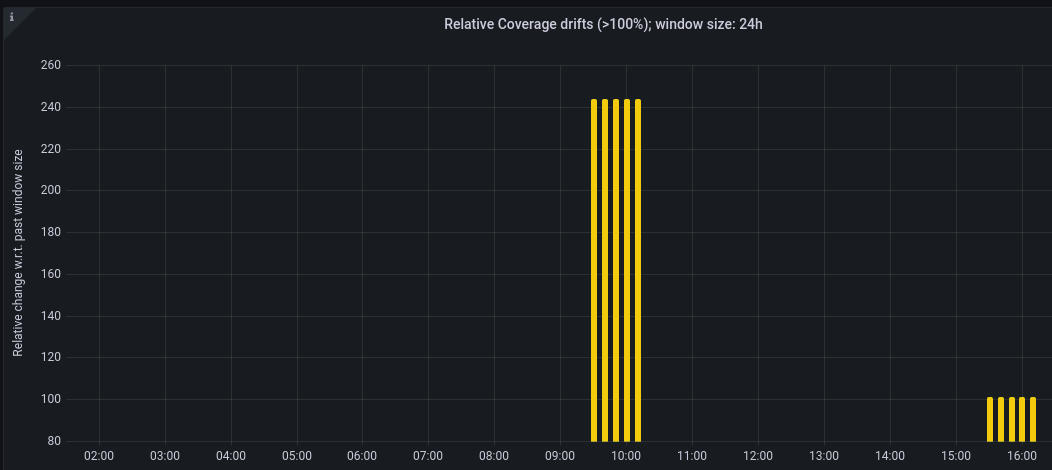}}
         \caption{Coverage drift events in real-time.}
     \end{subfigure}
        \caption{Different perspectives of a feature-level anomaly. Different metrics capture the anomaly, indicating a period where the feature was missing -- negative impact on its score is complementary to drops in coverage. The lift in standard deviation indicates that the process which generates the feature also changed, as even though its scores normalized after the anomaly, its deviation changed.}
        \label{fig:anomaly-example}
\end{figure*}
Models responsible for click-through and conversion rate prediction can be comprised of hundreds of features. By monitoring each feature that is in production's distribution and its shifts, Drifter instances alert the users when, e.g., a feature's coverage or cardinality score changes beyond expectations. We describe a use case where a feature that required effort from multiple teams to be productized required additional inspection as it misbehaved online -- being one of the more-relevant features, impact on model quality could be detected (with slight delay). The visualization layer of Drifter (Grafana-based visualizations of PromQL-based queries), apart from monitoring of the feature's distribution, also enables comparisons to previous time points. If the difference in a feature's coverage is beyond a pre-defined, acceptable threshold, Drifter logs it as an anomaly (visible in a designated dashboard panel), and can trigger the related alert.
Each visualization considers metrics derived from the main signals outputted by each Drifter instance. PromQL query examples are summarized in~Table~\ref{tbl:tblx}. The examples are straightforward to implement and are easily extensible with Prometheus's in-built capabilities related to computation of derivatives, linear extrapolation and similar. However, we observed that simple "deltas" between a quantity in a designated time frame already offer sufficient information for a better understanding of the issue at hand. Note that this setup does not require any external anomaly detection services (e.g., `prometheus-anomaly-detector`\footnote{\url{https://github.com/AICoE/prometheus-anomaly-detector}}), even though it can be extended to offer such functionalities should the need arise.
\begin{table*}[t!]
\caption{Overview of selected metrics that enable flexible monitoring of metric changes in time. The \emph{metric} part of specifications contains information about specific deployments, time intervals and ranges and other characteristics that enable detailed comparisons.}
\resizebox{\textwidth}{!}{
\begin{tabular}{l|l}
Metric's use                 & PromQL \\ \hline
Sliding differences - relative & $abs((\textrm{avg by} (\textrm{feature\_name}) (metric) \textrm{ offset interval})/ (\textrm{avg by} (\textrm{feature\_name}) (metric) * 100 > \textrm{threshold}$   \\
Sliding differences - absolute & $abs((\textrm{avg by} (\textrm{feature\_name}) (metric) \textrm{ offset interval}) - (\textrm{avg by} (\textrm{feature\_name}) (metric) * 100 > threshold$     \\
Outlier detection              & $\textrm{stddev by} (\textrm{feature\_name}) (metric) > 1/2 * \textrm{avg by} (\textrm{feature\_name}) (metric)$
\end{tabular}
}
\label{tbl:tblx}
\end{table*}

\section{Lessons learned and conclusions}
\label{sec-lessons}
Initially, Drifter was conceived as a stand-alone service with its front-end interface, enabling users to explore "personalized" visualizations independently of others. Even though part of this implementation remains available for each Drifter pod, we realized that the users' needs to define novel metrics of interest over longer time ranges could not be mimicked elegantly at the level of a custom in-house front-end solution. Furthermore, the architectural change that came with one UI per pod included extended data retention (per pod), causing additional disk overhead even though as soon as metrics are computed, raw outputs are no longer needed. By adopting the metric-push approach, we substantially reduced the computational resources required per Drifter pod and, at the same time, facilitated visualizations.
Drifter was initially designed to incorporate all required algorithms for fast and accurate ranking. However, by realizing that our in-house built, already optimized solution for feature ranking, enables fast-enough ranking out-of-the-box, we proceeded by creating an interface rather than writing a Drifter-specific solution. This way, algorithmic improvements already present in the proposed feature ranking engine could be further optimized and modified for Drifter. Furthermore, should the need arise, switching the ranking engine is straightforward, a matter of a changing interface.
We finally discuss the design choice of using the same data pipeline/processing steps as most machine learning flows. As Drifter can operate with raw Hive~\footnote{\url{https://hive.com}} dumps directly, there was initially no need to be entirely aligned with the processing steps undertaken to prepare data for, e.g., CTR prediction. However, we concluded that approximating the existing flows (data-wise) is a mandatory capability, as Drifters are, in most use cases, utilized to help explain anomalies or drifts of data that is fed to the prediction engine(s). This way, alignment up to the level of input (Vowpal Wabbit format - compressed) enabled us to approximate and enable the study of the very inputs that are used for downstream machine learning (e.g., CTR, CVR).
Finally, we discussed the design choices made along the way, to guide similar projects in avoiding the repetition of mistakes that were successfully addressed through the implementation of Drifter.
The feature ranking and verification engine used in Drifter will be made public soon as an open source project.

\bibliographystyle{ACM-Reference-Format}
\bibliography{sample-base}


\end{document}